\documentclass[aps,prd,twocolumn,nofootinbib,showpacs]{revtex4-1}

\usepackage{graphicx}

\usepackage[colorlinks=true, linkcolor=blue, citecolor=blue, urlcolor=blue]{hyperref}

\begin{document}

\title{The Generalized Scheme-Independent Crewther Relation in QCD}

\author{Jian-Ming Shen$^1$}
\email{cqusjm@cqu.edu.cn}
\author{Xing-Gang Wu$^1$}
\email{wuxg@cqu.edu.cn}
\author{Yang Ma$^2$}
\email{mayangluon@pitt.edu}
\author{Stanley J. Brodsky$^3$}
\email{sjbth@slac.stanford.edu}

\affiliation{$^1$Department of Physics, Chongqing University, Chongqing 401331, P.R. China}
\affiliation{$^2$Department of Physics and Astronomy, University of Pittsburgh, Pittsburgh, PA 15260, USA}
\affiliation{$^3$SLAC National Accelerator Laboratory, Stanford University, Stanford, California 94039, USA}

\begin{abstract}

The Principle of Maximal Conformality (PMC) provides a systematic way to set the renormalization scales order-by-order for any perturbative QCD calculable processes. The resulting predictions are independent of the choice of renormalization scheme, a requirement of renormalization group invariance. The Crewther relation, which was originally derived as a consequence of conformally invariant field theory, provides a remarkable connection between two observables when the $\beta$ function vanishes: one can show that the product of the Bjorken sum rule for spin-dependent deep inelastic lepton-nucleon scattering times the Adler function, defined from the cross section for electron-positron annihilation into hadrons, has no pQCD radiative corrections. The ``Generalized Crewther Relation" relates these two observables for physical QCD with nonzero $\beta$ function; specifically, it connects the non-singlet Adler function ($D^{\rm ns}$) to the Bjorken sum rule coefficient for polarized deep-inelastic electron scattering ($C_{\rm Bjp}$) at leading twist. A scheme-dependent $\Delta_{\rm CSB}$-term appears in the analysis in order to compensate for the conformal symmetry breaking (CSB) terms from perturbative QCD. In conventional analyses, this normally leads to unphysical dependence in both the choice of the renormalization scheme and the choice of the initial scale at any finite order. However, by applying PMC scale-setting, we can fix the scales of the QCD coupling unambiguously at every order of pQCD. The result is that both $D^{\rm ns}$ and the inverse coefficient $C_{\rm Bjp}^{-1}$ have identical pQCD coefficients, which also exactly match the coefficients of the corresponding conformal theory. Thus one obtains a new generalized Crewther relation for QCD which connects two effective charges, $\widehat{\alpha}_d(Q)= \sum_{i\geq1} \widehat{\alpha}^{i}_{g_1}(Q_i)$, at their respective physical scales. This identity is independent of the choice of the renormalization scheme at any finite order, and the dependence on the choice of the initial scale is negligible. Similar scale-fixed commensurate scale relations also connect other physical observables at their physical momentum scales, thus providing convention-independent, fundamental precision tests of QCD.

\end{abstract}



\maketitle

\section{Introduction}

The Crewther relation~\cite{Crewther:1972kn, Adler:1973kz} was originally derived as a remarkable consequence of conformally invariant field theory. It provides a nontrivial relation for three fundamental constants of a gauge theory with zero $\beta$ function and zero fermion masses:
\begin{equation}
3S=KR',  \label{eq:CCR}
\end{equation}
where $S$ is the anomalous constant governing $\pi_0\to\gamma\gamma$ decay~\cite{Adler:1969gk}, $K$ is the coefficient of the Bjorken sum rules for the polarized deep-inelastic electron scattering~\cite{Bjorken:1966jh}, and $R'$ is the isovector part of the cross-section ratio for the electron-positron annihilation into hadrons~\cite{Adler:1974gd}.

In QCD~\cite{Politzer:1973fx, Gross:1973id}, the logarithmic derivative of the running coupling, the $\beta$ function, is nonzero and negative; thus the strong coupling $\alpha_s$ becomes small at short distances due to asymptotic freedom, allowing the perturbative calculation of high-momentum transfer processes. Since physical QCD is non-conformal, the Crewther relation will be explicitly broken by the $\beta$-dependent terms arising from pQCD radiative corrections. One thus defines a ``Generalized  Crewther Relation (GCR)" by incorporating perturbative QCD (pQCD) approximants to each observable~\cite{Broadhurst:1993ru, Brodsky:1995tb, Crewther:1997ux}; i.e.,
\begin{eqnarray}
D^{\rm ns}(a_s) C_{\rm Bjp}(a_s) &=& 1+\Delta_{\rm csb},  \label{eq:GCR}
\end{eqnarray}
where $a_s = \alpha_s/\pi$, $D^{\rm ns}$ is the non-singlet Adler function, and $C_{\rm Bjp}$ is the coefficient in the Bjorken sum rule for polarized deep-inelastic electron scattering. The $\Delta_{\rm csb}$-term is introduced in order to compensate for the conformal symmetry breaking (CSB)-terms from both $D^{\rm ns}$ and $C_{\rm Bjp}(a_s)$. Using the renormalization group equation (RGE), it is suggested as the following form~\cite{Broadhurst:1993ru}
\begin{eqnarray}
\Delta_{\rm csb} &=& \frac{\beta(a_s)}{a_s}\sum_{i\geq1}{K^{\rm ns}_i}{a_s^i} = -\sum_{i\geq2}\sum_{k=1}^{i-1} K^{\rm ns}_k \beta_{i-1-k} a_s^i, \label{eq:csbbeta}
\end{eqnarray}
where the $\beta$ function, $\beta(a_s)=-\sum_{i\geq0}\beta_{i} a_s^{i+2}$; the coefficients $K^{\rm ns}_k$ are independent of the $\{\beta_i\}$. Thus $\Delta_{\rm csb}$, as constructed in Eq.(\ref{eq:csbbeta}), is a linear function of $\{\beta_i\}$-terms. The $\beta_0$ and $\beta_1$ terms are universal, whereas all high-order $\{\beta_{i\geq2}\}$-terms are scheme-dependent.

\section{Relations between physical observables and effective charges}

An essential requirement of the renormalization group is that relations between physical observables cannot depend on a theoretical convention such as the choices of the renormalization scheme and scale. This property is called ``renormalization group invariance" (RGI)~\cite{GellMann:1954fq, Bogolyubov:1956gh}. It is usually argued that the GCR only becomes renormalization scheme-and-scale independent at infinite order, leaving renormalization scheme-and-scale dependence at any finite order~\cite{Wu:2013ei}. For example, as will be shown later, the linear property of $\Delta_{\rm csb}$ as suggested by Eq.(\ref{eq:csbbeta}) persists only when both $D^{\rm ns}$ and $C_{\rm Bjp}$ are renormalized using the $\overline{\rm MS}$-scheme. The utility of the GCR (\ref{eq:GCR}) at finite order in pQCD is thus greatly constrained.

Does there exist a general scheme-independent GCR for pQCD theory at any fixed order? In fact, as we shall show in this article, a fundamental scheme-independent scale-fixed GCR can be achieved at each finite order by applying the ``Principle of Maximal Conformality" (PMC)~\cite{Brodsky:2011ta, Brodsky:2011ig, Mojaza:2012mf, Brodsky:2013vpa}.

The PMC is designed to eliminate renormalization scheme-and-scale ambiguities simultaneously. The PMC has a rigorous theoretical foundation and satisfies the essential properties of RGI~\cite{Brodsky:2012ms, Wu:2014iba}. It provides the underlying principle for the Brodsky-Lepage-Mackenzie (BLM) method~\cite{Brodsky:1982gc}, and it reduces to the standard Gell Mann-Low method~\cite{GellMann:1954fq} in the $N_C \to 0$ Abelian limit~\cite{Brodsky:1997jk}. As in QED, separate renormalization scales and effective numbers of quark flavors ($n_f$) appear for each skeleton graph, reflecting their differing virtualities at each perturbative order. The coefficients of the resulting pQCD series match the coefficients of the corresponding conformal theory with $\beta=0$, ensuring the scheme-independence of the PMC predictions at any fixed order. The usual divergent $n! \alpha_s^n \beta_0^n$ renormalon terms in the pQCD series do not appear. Furthermore, the PMC predictions have negligible residual scheme-and-scale dependence on the choice of the initial renormalization scale.

An important concept in QCD is the ``effective charge"~\cite{Grunberg:1980ja, Grunberg:1982fw} --  running couplings which can be defined from any perturbatively calculable observable. In the case of the GCR, we will introduce two effective couplings defined by the non-singlet Adler function $D^{\rm ns}$ and the Bjorken sum rules $C_{\rm Bjp}$,
\begin{eqnarray}
D^{\rm ns}(a_s) &=& 1 + \widehat{\alpha}_d(Q), \label{eq:aDns} \\
C_{\rm Bjp}(a_s) &=& 1 - \widehat{\alpha}_{g_1}(Q), \label{eq:ag1}
\end{eqnarray}
where $Q$ is the mass scale of the observable -- the kinematic value at which it is measured. Substituting Eqs.(\ref{eq:aDns}, \ref{eq:ag1}) into Eq.(\ref{eq:GCR}), we obtain
\begin{eqnarray}
[1+\widehat{\alpha}_d(Q)][1-\widehat{\alpha}_{g_1}(Q)] &=& 1+\Delta_{\rm csb}. \label{eff:GCR}
\end{eqnarray}

An effective coupling is by definition pQCD calculable; the $\beta$-pattern for its perturbative series at each order is a superposition of all of the $\{\beta_i\}$-terms which govern the evolution of the lower-order $\alpha_s$ contributions appearing at this particular order. More explicitly, the $\beta$-patterns for the pQCD approximants of these two effective couplings $\widehat{\alpha}_d(Q)$ and $\widehat{\alpha}_{g_1}(Q)$ are
\begin{eqnarray}
\widehat{\alpha}_{g_1} &=& r^C_{1,0}a_s + (r^C_{2,0} + \beta_0 r^C_{2,1} ) a_s^2  \nonumber\\
&& +(r^C_{3,0} + \beta_1 r^C_{2,1} + 2 \beta_0 r^C_{3,1} + \beta_0^2 r^C_{3,2} ) a^3_s \nonumber\\
&& +(r^C_{4,0} + \beta_2 r^C_{2,1}  + 2 \beta_1 r^C_{3,1} +  \frac{5}{2} \beta_1 \beta_0 r^C_{3,2}  \nonumber\\
&& +3 \beta_0 r^C_{4,1} + 3 \beta_0^2 r^C_{4,2} + \beta_0^3 r^C_{4,3} ) a_s^4+ \cdots
\label{eq:Cbetapattern}
\end{eqnarray}
and
\begin{eqnarray}
\widehat{\alpha}_d &=& r^D_{1,0}a_s + (r^D_{2,0} + \beta_0 r^D_{2,1} ) a_s^2  \nonumber\\
&& +(r^D_{3,0} + \beta_1 r^D_{2,1} + 2 \beta_0 r^D_{3,1} + \beta_0^2 r^D_{3,2} ) a^3_s \nonumber\\
&& +(r^D_{4,0} + \beta_2 r^D_{2,1}  + 2 \beta_1 r^D_{3,1} +  \frac{5}{2} \beta_1 \beta_0 r^D_{3,2}  \nonumber\\
&& +3 \beta_0 r^D_{4,1} + 3 \beta_0^2 r^D_{4,2} + \beta_0^3 r^D_{4,3} ) a_s^4+\cdots,
\label{eq:Dbetapattern}
\end{eqnarray}
where the $r^{C,D}_{i,j=0}$ are conformal coefficients and $r^{C,D}_{i,j\neq0}$ are non-conformal ones. Thus we obtain
\begin{widetext}
\begin{eqnarray}
\Delta_{\rm csb} &=& \left(r^D_{1,0}-r^C_{1,0}\right)a_s+\left[r^D_{2,0}-r^C_{1,0}r^D_{1,0}-r^C_{2,0} +(r^D_{2,1}-r^C_{2,1})\beta_0\right]a_s^2 +\bigg[r^D_{3,0}-r^C_{2,0}r^D_{1,0}-r^C_{1,0}r^D_{2,0}-r^C_{3,0}  \nonumber\\
&& +(2r^D_{3,1}-r^C_{2,1} r^D_{1,0}-r^C_{1,0} r^D_{2,1}-2 r^C_{3,1})\beta_0+(r^D_{2,1}-r^C_{2,1})\beta_1+(r^D_{3,2}-r^C_{3,2})\beta_0^2 \bigg]a_s^3 \nonumber\\
&& +\bigg[r^D_{4,0}-r^C_{3,0} r^D_{1,0}-r^C_{2,0} r^D_{2,0}-r^C_{1,0} r^D_{3,0}-r^C_{4,0}+\left(3r^D_{4,1}-2 r^C_{3,1} r^D_{1,0}-r^C_{2,1} r^D_{2,0}-r^C_{2,0} r^D_{2,1}-2 r^C_{1,0} r^D_{3,1}-3 r^C_{4,1}\right)\beta_0  \nonumber\\
&& +\left(-r^C_{2,1} r^D_{1,0}-r^C_{1,0} r^D_{2,1}-2 r^C_{3,1}+2 r^D_{3,1}\right)\beta_1+\left(r^D_{2,1}-r^C_{2,1}\right)\beta_2 +\frac{5}{2}(r^D_{3,2}-r^C_{3,2})\beta_1 \beta_0  \nonumber\\
&& +\left(3r^D_{4,2}-r^C_{3,2} r^D_{1,0}-r^C_{2,1} r^D_{2,1}-r^C_{1,0} r^D_{3,2}-3 r^C_{4,2}\right)\beta_0^2+(r^D_{4,3}-r^C_{4,3})\beta_0^3 \bigg]a_s^4+\mathcal{O}(a_s^5).
\label{eq:csbExpan}
\end{eqnarray}
\end{widetext}
The coefficients of $\beta_{i-2} a_s^i$ ($i\geq2$), the coefficients of $\beta_{i-3} a_s^i$ ($i\geq3$), etc., also follow from the PMC degeneracy relations among different orders~\cite{Mojaza:2012mf, Brodsky:2013vpa}, as shown explicitly by Eqs.(\ref{eq:Cbetapattern}, \ref{eq:Dbetapattern}). The degeneracy relations explain why one only needs to introduce one new parameter at each new order in the perturbative series (\ref{eq:csbbeta}) for $\Delta_{\rm csb}$.

The above formulae are general and applicable for any renormalization scheme. However, If one uses dimensional regularization, specifically, the $\overline{\rm MS}$-scheme, the form of $\Delta_{\rm csb}$ will be greatly simplified. For example, the conformal coefficients of $D^{\rm ns}$ and $C_{\rm Bjp}$ in the $\overline{\rm MS}$-scheme can be derived from the pQCD series derived in Refs.\cite{Chetyrkin:1994js, Baikov:2010je, Baikov:2012zm, Baikov:2012zn} which have been computed up to four-loop level. They satisfy the following relations
\begin{eqnarray}
r^D_{n,0}-\sum_{i=1,j=1}^{i+j=n}{r^C_{i,0} r^D_{j,0}}-r^C_{n,0}=0.\;\;\; (n\geq1)
\label{eq:confRelation}
\end{eqnarray}
There are also simple relations among the non-conformal coefficients when one uses the $\overline{\rm MS}$-scheme,
\begin{eqnarray}
&& r^D_{3,2}-r^C_{3,2} = 0, \;\;
r^D_{4,3}-r^C_{4,3} = 0, \\
&& 3r^D_{4,2}-r^C_{3,2} r^D_{1,0}-r^C_{2,1} r^D_{2,1}-r^C_{1,0} r^D_{3,2}-3 r^C_{4,2} = 0.
\end{eqnarray}
Comparing with Eq.(\ref{eq:csbbeta}), we obtain
\begin{eqnarray}
K^{\rm ns}_1 &=& r^C_{2,1}-r^D_{2,1},  \\
K^{\rm ns}_2 &=& 2r^C_{3,1}+r^C_{2,1} r^D_{1,0}+r^C_{1,0} r^D_{2,1}-2r^D_{3,1},  \\
K^{\rm ns}_3 &=& 3r^C_{4,1}+2r^C_{3,1} r^D_{1,0}+r^C_{2,1} r^D_{2,0}+r^C_{2,0} r^D_{2,1}  \nonumber\\
&& +2 r^C_{1,0} r^D_{3,1}-3r^D_{4,1}.
\end{eqnarray}
This confirms the previously observation of linear features of $\Delta_{\rm csb}$ using the $\overline{\rm MS}$-scheme at the next-to-next-to-leading order level~\cite{Broadhurst:1993ru}. It is the degeneracy relations among different orders that ensures the linear property of $\Delta_{\rm csb}$; one can then derive the coefficients $r^C_{i,j}$ and $r^D_{i,j}$ from the dependence of the $\beta_i$ on $n_f$ with the help of the degeneracy relations.

\section{Generalized anomalous dimensions}

It should be noted that the simplifying linear properties of $\Delta_{\rm csb}$ are not obtained in general. We will illustrate this by using the generalized dimensional regularization ${\cal R}_\delta$-scheme introduced in Ref.\cite{Brodsky:2013vpa}. In the ${\cal R}_\delta$-scheme one defines the UV regularization by subtracting an extra arbitrary constant $\delta$: $\ln 4\pi -\gamma_E- \delta$. The conventional $\overline{\rm MS}$-scheme corresponds to $\delta=0$. The resulting $\beta$-function is the same for all ${\cal R}_\delta$-schemes. Thus a scale transformation between different ${\cal R}_\delta$-schemes corresponds simply to a scale displacement, such as $\mu^2_\delta=\mu^2\exp(\delta)$~\cite{Mojaza:2012mf}, where $\mu$ stands for the scale of the $\overline{\rm MS}$-scheme. The displacement relation implies
\begin{equation} \label{displace}
a_s = a_{\delta} + \sum_{n=1}^\infty \frac{1}{n!} {\left. \frac{{\rm d}^n a_s}{({\rm d} \ln \mu^2)^n}\right|_{\delta} (-\delta)^n},
\end{equation}
where $a_s=a_s(\mu)$ and $a_\delta=a_s(\mu_\delta)$. The conformal-breaking term under the ${\cal R}_\delta$-scheme thus becomes
\begin{eqnarray}
\Delta_{\rm csb}[\delta] = -\sum_{i=2}^{\infty}\sum_{k=1}^{i-1} K^{\rm ns}_k[\delta] \beta^{\cal R}_{i-1-k} a_s^i, \label{d:eq:csbbeta}
\end{eqnarray}
in which the first three coefficients are
\begin{eqnarray}
K^{\rm ns}_1[\delta] &=& K^{\rm ns}_1,  \\
K^{\rm ns}_2[\delta] &=& K^{\rm ns}_2 + 2 \delta \beta_0 K^{\rm ns}_1,  \\
K^{\rm ns}_3[\delta] &=& K^{\rm ns}_3 + 3 \delta \beta_0 K^{\rm ns}_2 + 3 (\delta \beta_1 + \delta^2 \beta_0^2) K^{\rm ns}_1.
\end{eqnarray}
Here $K^{\rm ns}_i\equiv K^{\rm ns}_i[0]$ corresponds to the case of the $\overline{\rm MS}$-scheme. Obviously, for any other ${\cal R}_{\delta\neq0}$-scheme, the $\delta\neq0$-terms explicitly break the linear property of $\Delta_{\rm csb}$. In other words, the scheme-dependence of the fixed-order pQCD series breaks the generality of $\Delta_{\rm csb}$ and hence the generality of the GCR (\ref{eq:GCR}).

In contrast, as we shall show, the PMC sets the optimal scale for each pQCD approximant at each order by absorbing every $\beta$ term of the pQCD series into its respective running coupling, thus providing scheme-independent fixed-order pQCD predictions. In particular, we will obtain a generalized scheme-independent Crewther relation (GSICR) for QCD.

As an illustration, we will derive the PMC predictions for $D^{\rm ns}$ and $C_{\rm Bjp}^{-1}$. The non-singlet Adler function is defined as~\cite{Adler:1974gd},
\begin{eqnarray}
D^{\rm ns}(a_s) = -12{\pi^2}{Q^2}\frac{d}{d{Q^2}}\Pi^{\rm ns}(L,{a_s}),
\label{eq:Adler}
\end{eqnarray}
where $L=\ln{\mu^2}/{Q^2}$, $\mu$ encodes the renormalization scale. The coefficient $\Pi^{\rm ns}(L,{a_s})$ is the non-singlet part of the polarization function for a flavor-singlet vector current. The running behavior of $\Pi^{\rm ns}(L,a_s)$ is controlled by
\begin{eqnarray}
\left( {\mu^2}\frac{\partial}{\partial{\mu^2}} + \beta(a_s)\frac{\partial}{\partial{a_s}} \right)\Pi^{\rm ns}(L,a_s) &=& \gamma^{\rm ns}(a_s) ,
\label{eq:gamma}
\end{eqnarray}
where $\Pi^{\rm ns}(L,a_s)=\sum\limits_{i \ge 0} {\Pi_i^{\rm ns}} {a_s^i}/{16\pi^2}$. The Adler function depends on the anomalous dimension of the photon field $\gamma^{\rm ns}(a_s) = \sum\limits_{i \ge 0} {\gamma^{\rm ns}_i} {a_s^i}/{16\pi^2}$; its perturbative coefficients $\gamma^{\rm ns}_i$ using the $\overline{\rm MS}$-scheme up to four-loop level can be found in Ref.\cite{Baikov:2012zm}.

Thus we obtain
\begin{equation}
D^{\rm ns}(a_s) = 12\pi^2 \left[\gamma^{\rm ns}(a_s) - \beta(a_s)\frac{\partial} {\partial{a_s}}{\Pi}^{\rm ns}(L,{a_s})\right]. \label{eq:Dexpression}
\end{equation}
With the help of Eqs.(\ref{eq:Adler}, \ref{eq:gamma}), we observe $d D^{\rm ns}/d\mu^2=0$ at any fixed order, the pQCD approximant $D^{\rm ns}$ is thus a local RGI quantity.

It should be emphasized that the anomalous dimension $ \gamma^{\rm ns}(a_s)$ is associated with the renormalization of the QED coupling, it determines the correct running behavior of $\Pi^{\rm ns}(L,a_s)$; and it ensures that $D^{\rm ns}$ satisfy the local RGI, but not the standard RGI~\cite{Wu:2014iba}. This explains why the QED anomalous dimension $\gamma^{\rm ns}$, which appears in the definition of the Adler function, cannot be used to set the pQCD renormalization scales for $D^{\rm ns}$.

\section{The ratio ${D^{\rm ns}(a_s)}/{C_{\rm Bjp}^{-1}(a_s)}$ before and after PMC scale-setting}

\begin{figure}[htb]
\includegraphics[width=0.48\textwidth]{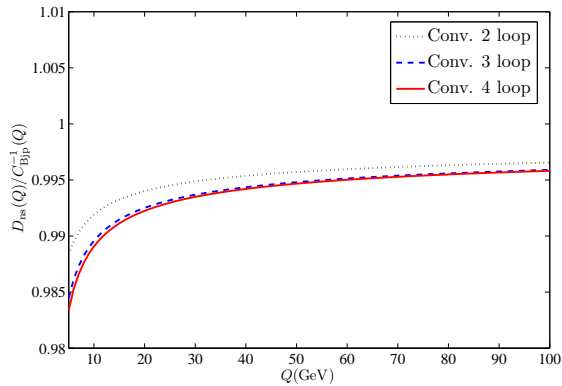}
\caption{Numerical predictions for the ratio ${D^{\rm ns}(a_s)}/{C_{\rm Bjp}^{-1}(a_s)}$ up to the four-loop level using conventional scale-setting.}
\label{fig0}
\end{figure}

The dependence of the ratio ${D^{\rm ns}(a_s)}/{C_{\rm Bjp}^{-1}(a_s)}$ on the momentum scale $Q$ which is predicted using conventional scale-setting is illustrated in Fig.(\ref{fig0}). One observes a small deviation of the ratio from unity, which cannot be diminished even by including more loop corrections. At $Q\sim 100$ GeV, the ratio is $\simeq 0.997$ for the two-loop
corrections, which changes to $\simeq 0.996$ for the three- and four-loop corrections. This small derivation occurs because the non-zero $\Delta_{csb} $ starts at $\alpha_s^2$ -order, as shown by Eq.(\ref{eq:csbbeta}). The scheme-dependent conformal-breaking term $\Delta_{\rm csb}$ accounts for the scheme-dependence of ${D^{\rm ns}(a_s)}/{C_{\rm Bjp}^{-1}(a_s)}$ as predicted by the GCR (\ref{eq:GCR}); it also explains why the ratio deviates from unity.

\begin{figure}[htb]
\includegraphics[width=0.48\textwidth]{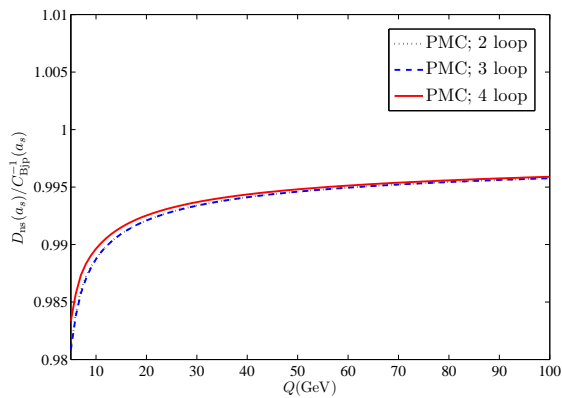}
\caption{Numerical predictions for the ratio ${D^{\rm ns}(a_s)}/{C_{\rm Bjp}^{-1}(a_s)}$ up to four-loop level after applying PMC scale-setting.}
\label{fig}
\end{figure}

\begin{figure}[htb]
\includegraphics[width=0.48\textwidth]{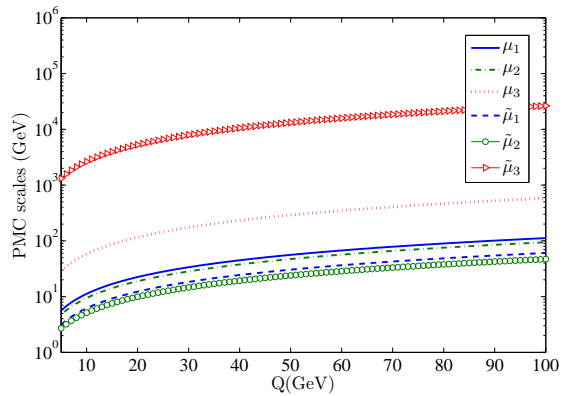}
\caption{The six PMC scales $\mu_{i}$ and $\tilde{\mu}_i$ versus the kinematic scale $Q$ of $D^{\rm ns}(a_s)$ and $C^{-1}_{\rm Bjp}(a_s)$.}
\label{figscale}
\end{figure}

The scheme dependence of the generalized Crewther relation is eliminated when one applies PMC scale-setting. The PMC utilizes the RGE recursively to identify the $\beta$-pattern of any pQCD expansion. Eq.(\ref{eq:Dexpression}) shows that, except for the anomalous coefficient $\gamma^{\rm ns}$, which is associated with the renormalization of $\alpha_{\rm QED}$, the only remaining terms in $D^{\rm ns}$ are the pQCD RGE $\beta_i$-terms. After shifting the nonconformal $\beta_i$ terms into the scales of QCD coupling, one recovers the conformal series and determines the optimal scales of the running coupling at each order.

Thus, after applying the PMC ~\cite{Brodsky:2013vpa}, we obtain
\begin{equation}
D^{\rm ns}(a_s)|_{\rm{PMC}}= 12\pi^2 \gamma^{\rm ns}(\mu_1,\mu_2,\mu_3,\cdots).
\label{eq:nsAdlerPMC}
\end{equation}
Similarly, after applying PMC scale-setting to $C_{\rm Bjp}^{-1}(a_s)$, we obtain
\begin{equation}
C_{\rm Bjp}^{-1}(a_s)|_{\rm PMC} = 12\pi^2 \gamma^{\rm ns}(\tilde{\mu}_1, \tilde{\mu}_2, \tilde{\mu}_3, \cdots).  \label{eq:BjpPMC}
\end{equation}
Here $C^{-1}_{\rm Bjp}$ is introduced for convenience. It has the same scheme-independent conformal coefficients as those of $D^{\rm ns}$, as ensured by the relations (\ref{eq:confRelation}).

The first three PMC scales $\mu_{1,2,3}$ or $\tilde{\mu}_{1,2,3}$ can be derived from the pQCD series for $D^{\rm ns}$ and $C_{\rm Bjp}^{-1}$ given in Refs.\cite{Baikov:2012zm, Baikov:2010je}. The dependence of the PMC scales on $Q$ are presented in Fig.(\ref{figscale}). The conformal coefficients of the scheme-independent $D^{\rm ns}(a_s)|_{\rm{PMC}}$ and $C_{\rm Bjp}^{-1}(a_s)|_{\rm PMC}$ are identical, but their PMC scales differ: $\mu_{i}\neq \tilde{\mu}_{i}$; thus, as is the case for the conventional prediction shown in Fig.(\ref{fig0}), one also has a small deviation from unity, as shown in Fig.(\ref{fig}):
\begin{equation}
\frac{D^{\rm ns}(a_s)|_{\rm PMC}}{C_{\rm Bjp}^{-1}(a_s)|_{\rm PMC}} \approx 1. \label{eq:GCRPMC0}
\end{equation}
This result shows that by applying PMC scale-setting, one can systematically improve the pQCD predictions for $D^{\rm ns}(a_s)$ and $C_{\rm Bjp}^{-1}(a_s)$, although, due to the conformal breaking of QCD, they cannot be exactly equal.

We shall demonstrate in the next section that a new scheme-independent GCR can be achieved by using the effective coupling approach.

\section{The scheme-independent generalized Crewther relation}

Following Eqs.(\ref{eq:Cbetapattern}, \ref{eq:Dbetapattern}), the effective coupling $\widehat{\alpha}_d(Q)$ can be expanded in terms of $\widehat{\alpha}_{g_1}(Q)$,
\begin{eqnarray}
\widehat{\alpha}_d(Q) &=& \widehat{\alpha}_{g_1}(Q) + (r_{2,0} + \beta_0 r_{2,1})\widehat{\alpha}_{g_1}^2(Q)+ (r_{3,0} + \nonumber\\
&& \beta_1 r_{2,1} + 2\beta_0 r_{3,1} + \beta_0^2 r_{3,2}) \widehat{\alpha}_{g_1}^3(Q) + \nonumber\\
&& (r_{4,0} + \beta_2^{g_1} r_{2,1} + 2\beta_1 r_{3,1} + \frac{5}{2} \beta_0 \beta_1 r_{3,2} + \nonumber\\
&& 3\beta_0 r_{4,1} + 3\beta_0^2 r_{4,2} + \beta_0^3 r_{4,3}) \widehat{\alpha}_{g_1}^4(Q) +\cdots.  \label{eq:GCRbeta}
\end{eqnarray}
Here the $\{\beta^{g_1}_i\}$ are the $\beta$-terms for the ${g_1}$-scheme, which can be related to the $\overline{\rm MS}$ coefficients by using $\beta^{g_1}=\frac{\partial \alpha_{g_1}}{\partial \alpha_{\overline{\rm MS}}}{\beta^{\overline{\rm MS}}}$. The coefficients $r_{i,j}$ are functions of $r^{C/D}_{i,j}$, whose expressions are given in Refs.\cite{Brodsky:2013vpa} together with the conformal coefficients $r_{i,0}\equiv1 \;(i=1,2,\cdots)$.

After applying PMC scale-setting, we obtain a new scheme-independent GCR (GSICR) up to $n_{\rm th}$-order,
\begin{eqnarray}
\widehat{\alpha}_d(Q)= \sum_{i=1}^{n}\widehat{\alpha}^{i}_{g_1}(Q_{i}). \label{eqGCR}
\end{eqnarray}
At the known four-loop level ($n=4$), the first three PMC scales are
\begin{widetext}
\begin{eqnarray}
\ln\frac{Q_1^2}{Q^2} &=& -r_{2,1}+\widehat{\alpha}_{g_1}(Q)\beta_0({r^2_{2,1}}-r_{3,2}) +\widehat{\alpha}^2_{g_1}(Q)\bigg[\frac{3}{2}\beta_1(r^2_{2,1}-r_{3,2})+\beta_0^2(-{r^3_{2,1}} +2r_{2,1}r_{3,2}-r_{4,3}) \bigg]+\mathcal{O}(\widehat{\alpha}_{g_1}^3), \nonumber\\
\ln\frac{Q_2^2}{Q^2} &=& -\frac{r_{3,1}}{r_{2,0}}+\widehat{\alpha}_{g_1}(Q)\beta_0\left(\frac{3r_{4,2}}{2r_{2,0}} -\frac{3{r_{3,1}}^2}{2{r_{2,0}}^2}\right)+\mathcal{O}(\widehat{\alpha}_{g_1}^2),  \nonumber\\
\ln\frac{Q_3^2}{Q^2} &=& -\frac{r_{4,1}}{r_{3,0}}+\mathcal{O}(\widehat{\alpha}_{g_1}).
\end{eqnarray}
\end{widetext}

Eq.(\ref{eqGCR}) shows that the relation between $\widehat{\alpha}_d$ and $\widehat{\alpha}_{g_1}$ is independent of the renormalization scheme used to renormalize them. Together with the optimal scales set by the RGE, the GSICR is both scheme-and-scale independent at any fixed order. Such scale-fixed relations between physical observables are also called ``Commensurate Scale Relations" (CSRs)~\cite{Brodsky:1994eh}~\footnote{The CSRs given in Ref.\cite{Brodsky:1994eh} agree with ours in form. The relation (\ref{eq:confRelation}) (\ref{eqGCR})  also holds. However, these CSRs are derived without determining whether the $n_f$-terms at high orders pertain to the RGE or not; thus one cannot be certain that one obtains the correct running behavior at each order.}. In the conformally invariant limit where all $\{\beta_i\}$-terms tend to zero, we have $Q_i\equiv Q$ and we recover the scheme-independent conformal Crewther relation~\cite{Broadhurst:1993ru}
\begin{equation}
[1+\widehat{\alpha}_d(Q)][1-\widehat{\alpha}_{g_1}(Q)]=1.
\end{equation}

In conventional analyses, the $\Delta_{\rm csb}$-term is introduced into the original GCR, Eq.(\ref{eq:GCR}), to collect all of pQCD conformal breaking terms. The resulting $\Delta_{\rm csb}$-term is scheme dependent and only in the $\overline{\rm MS}$-scheme is it a linear expansion of $\{\beta_i\}$-functions.

However, by applying the PMC, we achieve a new fundamental scheme-independent GSICR, Eq.(\ref{eqGCR}), which is scheme-and-scale independent at any fixed order.

\section{Conclusion}

The PMC provides a systematic method to set the renormalization scale of high-energy processes, thus solving the conventional renormalization scheme-and-scale ambiguities. There is negligible dependence in the choice of the initial renormalization scale.

The GSICR given in Eq.(\ref{eqGCR}) provides a remarkable direct connection between deep inelastic neutrino-nucleon scattering and hadronic $e^+e^-$ annihilation. Similar scheme-independent GCRs can also be derived for many other physical observables, such as:

I) the connection of the Adler function $D=1+\widehat{\alpha}_D(Q)$ to the Gross-Llewellyn Smith (GLS) sum rule $C_{\rm GLS}=1-\widehat{\alpha}_{F_3}(Q)$~\cite{Gross:1969jf},
\begin{equation}
\widehat{\alpha}_{D}(Q) = \sum_{i\geq1} \widehat{\alpha}_{F_3}^i(\tilde{Q}_{i}).
\end{equation}

and

II) the CSR connection between the GLS sum rule and the $e^+e^-$ annihilation ratio $R_{e^+e^-}(Q)=3\sum q_f^2(1+\widehat{\alpha}_R(Q))$,
\begin{eqnarray}
\widehat{\alpha}_{F_3}(Q) = \sum_{i\geq1} (-1)^{i+1} \widehat{\alpha}_{R}^i(\tilde{\tilde{Q}}_{i}).
\end{eqnarray}
The PMC scales $\tilde{Q}_{i}$ and $\tilde{\tilde{Q}}$ can be determined using the standard PMC scale-setting procedure.

As is the case of the scheme-independent generalized Crewther relation, these scheme-and-scale independent commensurate scale relations provide fundamental, high precision tests of nonconformal QCD. For example, as an application of the GSICR, one may obtain a precise relation between $\alpha_{s,g_1}$ and $\alpha_{s,\overline{\rm MS}}$ which can be used to determine a well-defined effective charge $\alpha_{g_{1}}$ without scheme-dependence and with significantly improved precision~\cite{workinprogress}.  \\

\noindent{\bf Acknowledgement}: This work was supported in part by the National Natural Science Foundation of China under Grant No.11625520 and No.11275280, and the Department of Energy Contract No.DE-AC02-76SF00515. SLAC-PUB-16845. PITT-PACC-1612.

\appendix

\section*{Appendix: comparison of the ``two-fold" perturbation with the PMC}

We end with a  comment on the  use of the ``two-fold" perturbative expansion, which has been suggested for determining the $\beta$-pattern of a pQCD approximant~\cite{Kataev:2010du, Cvetic:2016rot}. Using this procedure, a physical observable $\rho$ can be approximated by the perturbative expansion
\begin{eqnarray}
\rho &=& \sum_{n\geq0,m\geq1} \bigg(\frac{\beta(a_s)}{a_s}\bigg)^n p_{m+n,n}a_s^m \nonumber\\
&=& p_{1,0} a_s(\mu) + (p_{2,0} + \beta_0 p_{2,1} ) a_s^2(\mu) +  \nonumber\\
&& (p_{3,0} + \beta_1 p_{2,1} + \beta_0 p_{3,1} + \beta_0^2 p_{3,2} ) a_s^3(\mu) + \nonumber\\
&&(p_{4,0} + \beta_2 p_{2,1}  + \beta_1 p_{3,1} +  2 \beta_1 \beta_0 p_{3,2} + \nonumber\\
&& \beta_0 p_{4,1} + \beta_0^2 p_{4,2} + \beta_0^3 p_{4,3} ) a_s^4(\mu)  + {\cal O}(a_s^5),
\label{eq:twofoldbeta}
\end{eqnarray}
where the coefficients $p_{m+n,n}$ are free from $\{\beta_i\}$-terms. The two-fold perturbation expansion generates the same $\beta$-pattern as that of PMC $R_{\delta}$-scheme at each order~\cite{Ma:2015dxa}. However, in comparison with the PMC expansion, such as Eqs.(\ref{eq:Cbetapattern}, \ref{eq:Dbetapattern}), the degeneracy relations introduced by the two-fold perturbation expansion is different from the PMC ones, even at order $a_s^4$.

The degeneracy relations introduced by the PMC are required by the conformality of the final series~\cite{Bi:2015wea}; they show that the $\beta$-pattern for the pQCD series at each order is a superposition of the $\{\beta_i\}$-terms which govern the evolution of all of the lower-order $\alpha_s$ contributions. Conversely, they determine the correct running behavior of $\alpha_s$ at each order.

In contrast, the $\beta$-pattern of the two-fold perturbation expansion can only yield an approximate running behavior at the contributing order; in effect it is equivalent to using the following approximate solution of the RGE to set the scale of the running coupling:
\begin{eqnarray}
a_s(\mu) = \frac{a_s(\mu_0)}{1- {\beta(a_s(\mu_0))}/{a_s(\mu_0)} \ln\frac{\mu^2}{\mu^2_0}},
\end{eqnarray}
where $\mu_0$ is some initial scale. Using the scale displacement equation, one obtains the $\beta$-pattern and degeneracy relations of Eq.(\ref{eq:twofoldbeta}).

\end{document}